\documentclass{aa}
\usepackage{graphicx}
\begin{document}
   \title{Optical and near infrared observations of SN\,1998bu
   \thanks{Based on observations collected at the La Silla and Paranal
   observatories of the European Southern
   observatory, Chile in time allocated to proposals 63.H-0527, 63.H-0649.
   }}


   \author{J. Spyromilio\inst{1}, 
          R. Gilmozzi\inst{1}, J. Sollerman\inst{2}, B. Leibundgut\inst{1},
          C. Fransson\inst{2} and J-G. Cuby\inst{3}}

   \offprints{J. Spyromilio}

   \institute{European Southern Observatory,
              Karl-Schwarzschild-Strasse 2, D-85748 Garching, Germany \\
              \email{jspyromi@eso.org, rgilmozz@eso.org,
              bleibundgut@eso.org}
              \and
              Stockholm Observatory, Albanova, SE-106 91 Stockholm,
              Sweden \\
              \email{
              jesper@astro.su.se,
              claes@astro.su.se}
	      \and
              Laboratoire d'Astrophysique de Marseille, B.P.8,
	      F-13379, Marseille, Cedex 12, France\\
	      \email{jean-gabriel.cuby@oamp.fr}
             }

   \date{Received ; accepted }

   \abstract{Infrared and optical spectra of SN\,1998bu at an age of
   one year after explosion are presented. The data show evidence for
   the radioactive decay of $^{56}$Co to $^{56}$Fe, long assumed to be
   the powering source for the supernova light curve past maximum
   light. The spectra provide direct evidence for at least 
   0.4 solar masses of
   iron being present in the ejecta of the supernova. The fits to the
   data also show that the widths of the emission lines increase with
   time. Photometric measurements in the H-band show that the supernova is not
   fading during the observation period. This is consistent with
   theoretical expectations. 
   }

\authorrunning{J. Spyromilio et al.}
\titlerunning{Supernova 1998bu}

\maketitle
%

\section{Introduction}

Type Ia supernovae are believed to be the most prodigious producers of
iron group elements and given the explosive nature of the events also
the best recyclers of such material. The burning of a large
fraction of a white dwarf near the Chandrasekhar limit into $^{56}$Ni
provides the energy for the explosion and subsequent radioactive decay of the
nickel to cobalt and then to stable iron provides the energy for the
light curve. Type Ia supernovae are popular distance indicators as
they exhibit, following a simple calibration procedure, remarkable
uniformity (see Phillips et al. \cite{phillips}). However, our
understanding of these objects remains limited (Leibundgut
\cite{leibundgut}).

The observation of the radioactive decay and accurate direct
measurements of the mass of $^{56}$Ni manufactured in the explosion provides
for a better underpinning of the theoretical scenario outlined
above. Evidence for radioactive decay comes from a number of
sources. The light curve shape foremost suggests an exponentially
declining power source.
Kuchner et al. (\cite{kuchner}) showed that
the flux ratio of a group of [Fe\,{\small III}] lines to [Co\,{\small
III}] increased with time in accordance with the expectations of
radioactive decay.  Sch{\"o}nfelder et al. (\cite{schoenfelder})
report on what would be the most direct and convincing evidence for
radioactive decay, namely the direct detection of the $\gamma$-ray
lines originating from the decay. However, even for the very luminous
SN\,1991T the detections are only marginal, although consistent with
the current theories. For SN\,1998bu Georgii et al. (\cite{georgii})
report upper limits for the amount of $^{56}$Ni manufactured in the explosion
from the absence of $^{56}$Co $\gamma$-ray emission. Their 2-$\sigma$
upper limit is 0.35\,M$_\odot$ of $^{56}$Ni.

The determination of the mass of $^{56}$Ni manufactured in the
explosion based on optical and infrared spectroscopy has
been the subject of many papers and almost as many
models. In this paper we present data collected in three observing
runs one year after the explosion of the supernova 1998bu. Two epochs of
infrared (IR) data are used to explore the evolution of the radioactive
element abundances. The later epoch is combined with optical data to
derive some limits on the mass of iron present in the ejecta of SN\,1998bu.

\section{Observations}

Supernova 1998bu was discovered in NGC 3368 (Villi \cite{villi}) on
May 9, 1998 and
reached maximum light in the $B$-band on May 21st 1998. The supernova
has been observed extensively by a number of groups (Suntzeff et
al. \cite{suntzeff}; Jha et al. \cite{jha}; Hernandez et
al. \cite{hernandez}). NGC~3368 (M96) has a distance modulus of
30.25$\pm$0.19 measured using Cepheid variables (Tanvir, Ferguson \&
Shanks \cite{tanvir}) and a recession velocity of 900$\pm$50\,km~s$^{-1}$. The
extinction towards the supernova is high with an $A_V = 0.94\pm0.15$
(Jha et al. \cite{jha}, see also Hernandez et al. \cite{hernandez} 
 who derive an $A_V$ of 1.0$\pm$0.1).

SN\,1998bu was observed with the SofI infrared camera spectrograph at
the 3.58-m ESO New Technology telescope at La Silla on January 26,
1999 at an age of 250 days (past maximum blue light).  Two grism
settings were used providing a spectrum covering the J, H and K bands
(Figure\,1). The spectroscopic standard BS4903 was used to remove
telluric features. The spectral resolution of the data is ($\lambda
/\delta\lambda$)=500. The absolute fluxing of the data was based on
the photometry obtained on the same night with the same instrument in
the $H$ band. The photometric standards were P550C and S217D (Persson
et al. \cite{persson}).  The supernova $H$-band magnitude was
17.5$\pm$0.1.

\begin{figure}
\centering
\includegraphics[angle=-90,width=8cm]{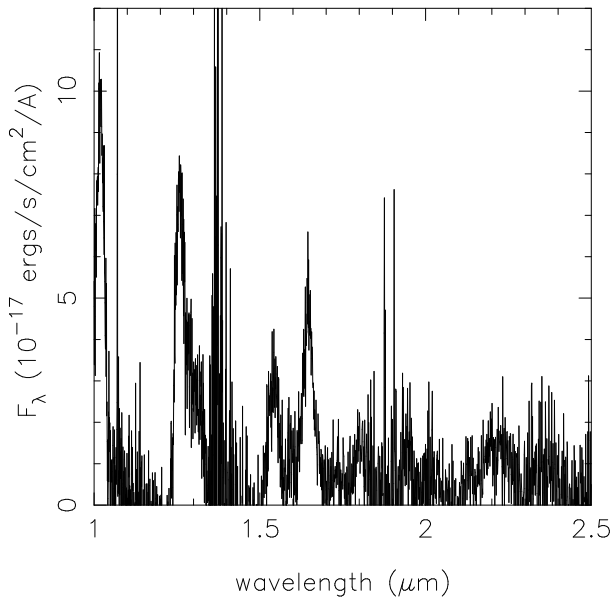}
 \caption{Spectrum of SN\,1998bu obtained with SofI at the NTT on
January 26, 1999.}
\label{sofi}
\end{figure}

\begin{figure}
\centering
\includegraphics[angle=-90,width=8cm]{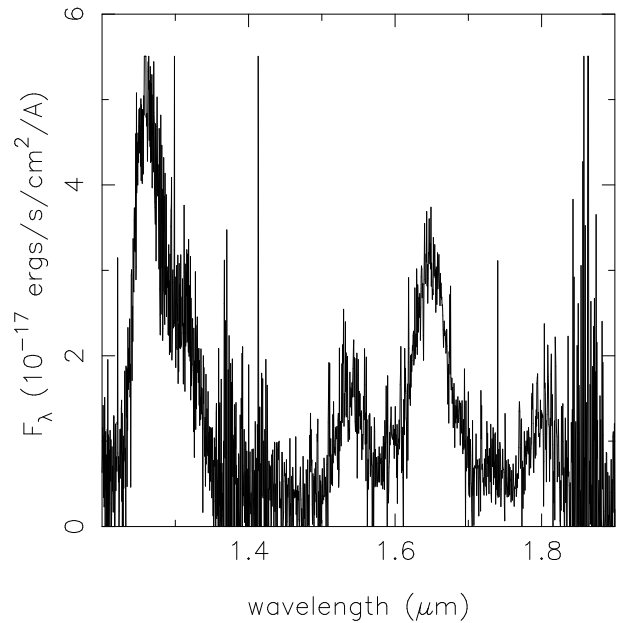}
 \caption{Spectrum of SN\,1998bu obtained with ISAAC at UT1 on
May 1, 1999.}
\label{isaac}
\end{figure}

The ISAAC infrared camera spectrograph at the 8.2\,-\,m ESO UT1 telescope
of the VLT array at Paranal was used to observe SN\,1998bu on May 1,
1999 at an age of 344 days.
Imaging observations were obtained with ISAAC to photometrically
calibrate the spectrum. The photometric standard was S860D (Persson et
al.  \cite{persson}) and the derived $H$-band magnitude was
17.5$\pm$0.1. The spectrum obtained covers the 1.2--1.9\,-$\mu$m
wavelength range (Figure\,2). The spectral resolution of the data is ($\lambda
/\delta\lambda$)=1500. The observations were obtained in two wavelength
settings and the spectra were joined without any scaling.  The
accuracy of merging spectra obtained in different settings without scaling
can be confirmed during the fitting of the
data. The 1.644-$\mu$m and 1.257-$\mu$m lines of
[Fe\,{\small II}] originate from the same upper level and their ratio is 
fixed by the radiative
transition rates and the extinction only. As will be shown below the
fits are excellent adopting standard atomic data and the commonly assumed
extinction for SN\,1998bu. 

FORS1 observations of SN\,1998bu at the 8.2-m ESO UT1 telescope of the
VLT array were obtained on June 6, 1999 at an age of 381 days (past
maximum blue light). An optical spectrum covering the
0.4--0.9\,-$\mu$m wavelength range was obtained.  The spectral
resolution of the data is ($\lambda /\delta\lambda$)=440.
The spectrophotometric standard EG274 was used to flux the
spectra.
The absolute flux of the spectrum was also checked against the
photometry derived from the acquisition images.  The night was
non-photometric and we have used the local standards as established by
Hernandez et al. (\cite{hernandez}) (stars 2, 3, 4 and 5) as well as
star 6 of Jha et al. (\cite{jha}) to derive the SN magnitude. The
scatter in the 5 local standards was less than 0.05 magnitudes and no
colour terms were applied.  The $V$-band magnitude of the supernova
was 19.96$\pm$0.10.

\section{Discussion}

\subsection{The $H$-band light curve}

The observational prescription for IR light curves of type Ia
supernovae is that they decline by 0.013 magnitudes per day and comes
from the work of Elias \& Frogel (\cite{elias}) and the more recent
update by Meikle (\cite{meikle}). However, these works do not extend
their coverage significantly into the first year after explosion.
Extrapolating this rule, supernova 1998bu when observed with ISAAC should
have been 1.2 magnitudes fainter than observed with SofI. This is not 
consistent with our observations (see Figure\,3).

To double-check our zero-points and the accuracy of the photometry we
performed photometry of 26 stars common in the field of view for the
two instruments. The chosen stars all had magnitudes within $\pm1.5$
mag of the supernova, and were not significantly contaminated by the
host galaxy. For the stars of comparable brightness to the supernova
we derive the same magnitude in both the ISAAC and SofI data with a
scatter consistent with the photometric errors in each
observation. There is no systematic offset between the two imaging
data sets and it is clear that within the errors of the photometry the
supernova has not faded between the two observations ($\pm$0.1
magnitude).

\begin{figure}
\centering \includegraphics[angle=-90,width=8cm]{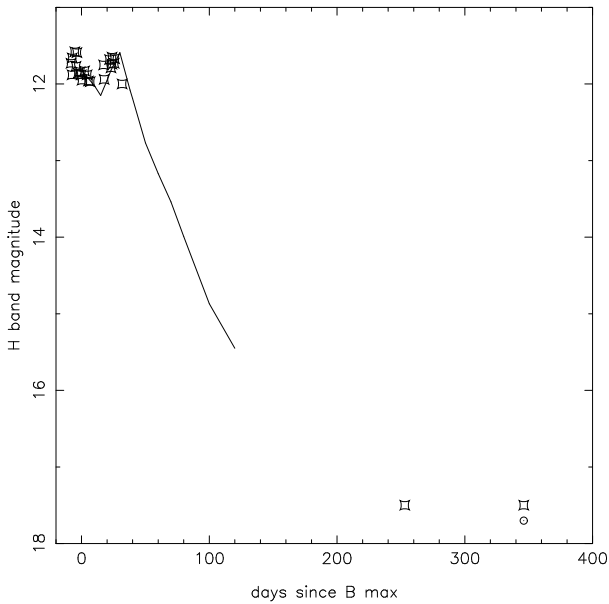}
 \caption{Light curve of SN\,1998bu in the $H$-band. Early data from
Meikle (\cite{meikle}) and template light curve from Elias
(\cite{elias}). For the observation at the epoch of 350 days the lower
point (dotted circle) indicates the magnitude of the supernova after
the removal of a possible contribution from continuum not arising from
the ejecta.}
\label{lightcurve}
\end{figure}

The possibility that a fraction of the flux is coming from an echo
(reflected light from earlier times that is contributing to the
observed magnitude of the supernova) is examined below. However, the
late emission line spectrum is not consistent with a significant
contribution from an echo.

\subsection{Removing the echo}

\begin{figure*}
\centering \includegraphics[angle=-90,width=17cm]{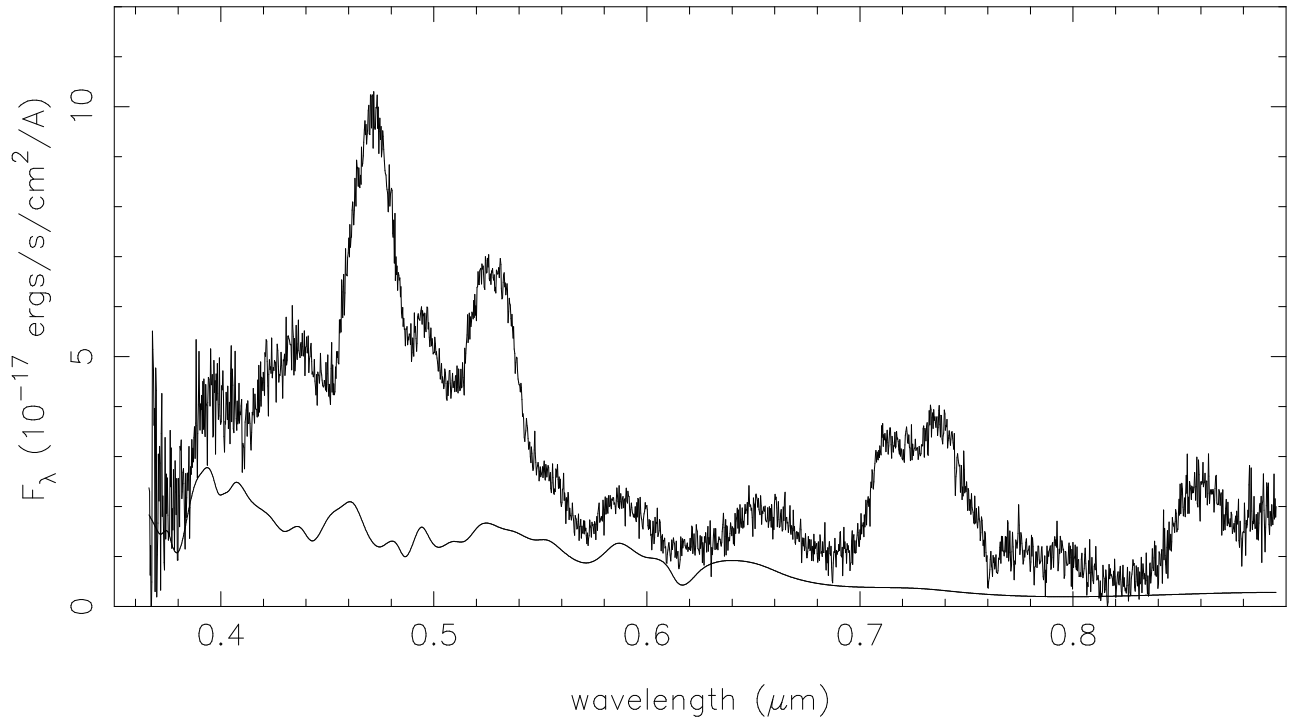}
 \caption{Spectrum of SN\,1998bu obtained with FORS1 at UT1 on June
11, 1999. The lower spectrum is a trace of the late light echo spectrum
(Cappellaro et al. \cite{cappellaro}) scaled by a factor of 2 upwards
- see text for details}
\label{fors}
\end{figure*}

Cappellaro et al. (\cite{cappellaro}) have demonstrated that the
optical spectrum of SN\,1998bu at an age of 600 days is dominated by
an echo. HST observations
have resolved a ring surrounding the supernova which is consistent
with emission from an echo.  Garnavich et al. (\cite{garnavich}) from
HST data determine the magnitude of the echo to be 21.4 in the visible
(a factor of 2 higher than that determined by Cappellaro et al.).
We traced the echo spectrum from Cappellaro et al. and 
have smoothed and scaled it by the HST photometry.
It is reproduced together with our FORS1 spectrum
in Fig.~\ref{fors}. At the epoch of the FORS1 spectrum no
significant continuum emission is expected from the supernova
ejecta. The scaled spectrum fits well under the FORS1 data and in all
fits to the data we have removed this contribution from our data.

The SofI data, obtained at a time when no contribution from the echo
was seen in the Cappellaro light curve, also show no evidence of a
continuum. The ISAAC data, however, do show evidence for an underlying
continuum at a level of
3$\times$10$^{-18}$\,erg~s$^{-1}$~cm$^{-2}$\AA$^{-1}$.  Without a
detailed model of the dust and its scattering properties it is
difficult to establish the exact contribution to this continuum by the
echo.  At the epochs under consideration in this paper, there is no
evidence from other supernovae for a continuum due to emission from
the ejecta. A level of
3$\times$10$^{-18}$\,erg~s$^{-1}$~cm$^{-2}$\AA$^{-1}$ is somewhat
higher than a simple extrapolation of the echo spectrum scaled as
mentioned above. However, given the broader $H$-band light curve it is
reasonable to expect the IR echo to be somewhat brighter than the optical 
one. We adopt this continuum level as a non-ejecta contribution.

Adjusting the supernova magnitude for this continuum results in a
H-band magnitude of 17.7 for the supernova ejecta at the epoch of the ISAAC
observations. We adopt the continuum subtracted spectrum as the one
originating from the ejecta. Our conclusions are only marginally
affected if we have overestimated the contribution from the echo
in the IR spectrum.

\subsection{Fitting the spectra}

\begin{figure*}
\centering \includegraphics[angle=-90,width=17cm]{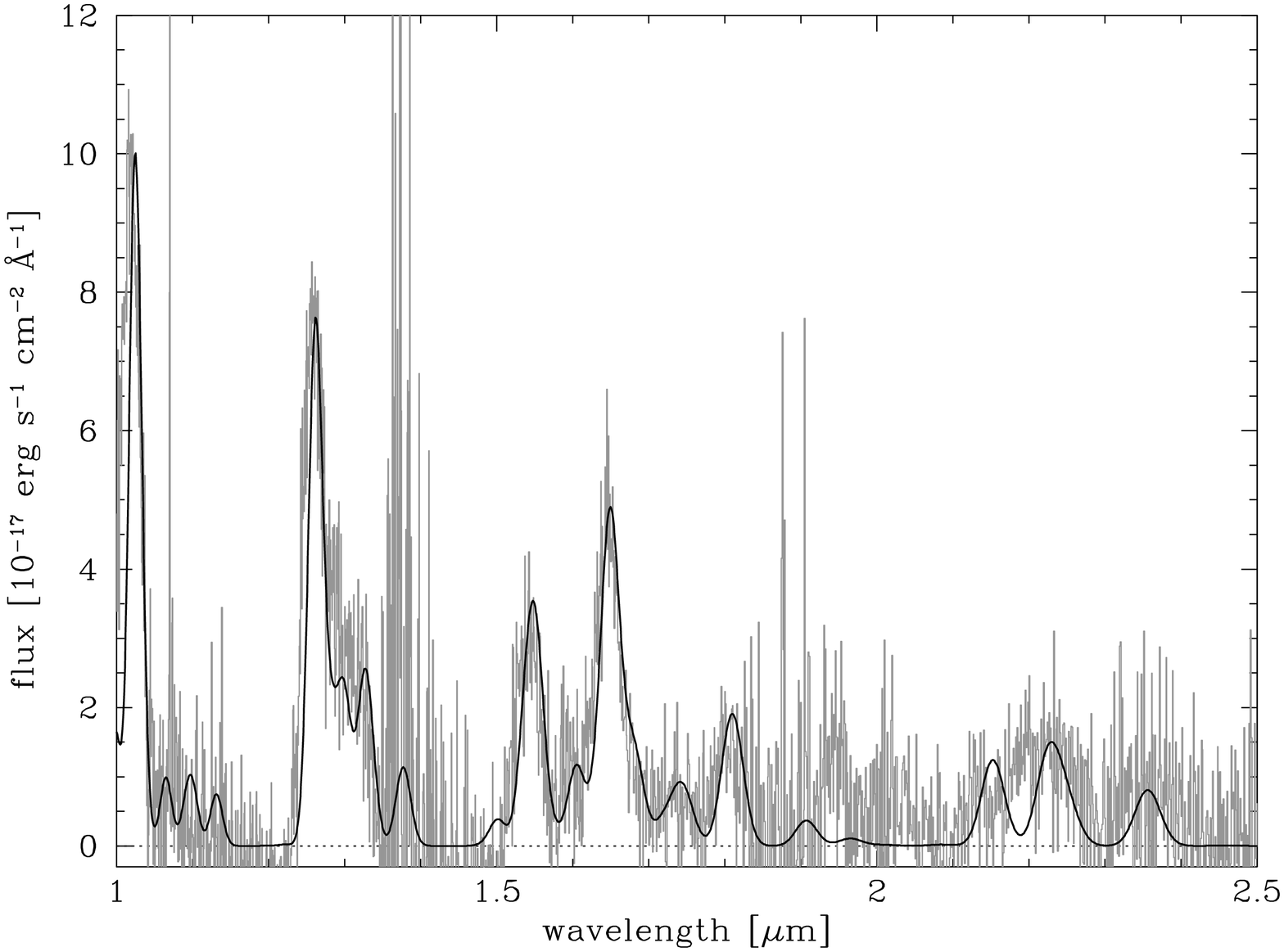}
 \caption{A fit to the spectrum obtained with SOFI at the NTT using a
simple NLTE [Fe\,{\small I}],  [Fe\,{\small II}] and [Fe\,{\small III}]
model and LTE [Co\,{\small II}]
The narrow (few pixel) spikes in the data are artifacts of
the atmosphere and/or the detector.}
\label{fitsofi}
\end{figure*}

\begin{figure*}
\centering \includegraphics[angle=-90,width=17cm]{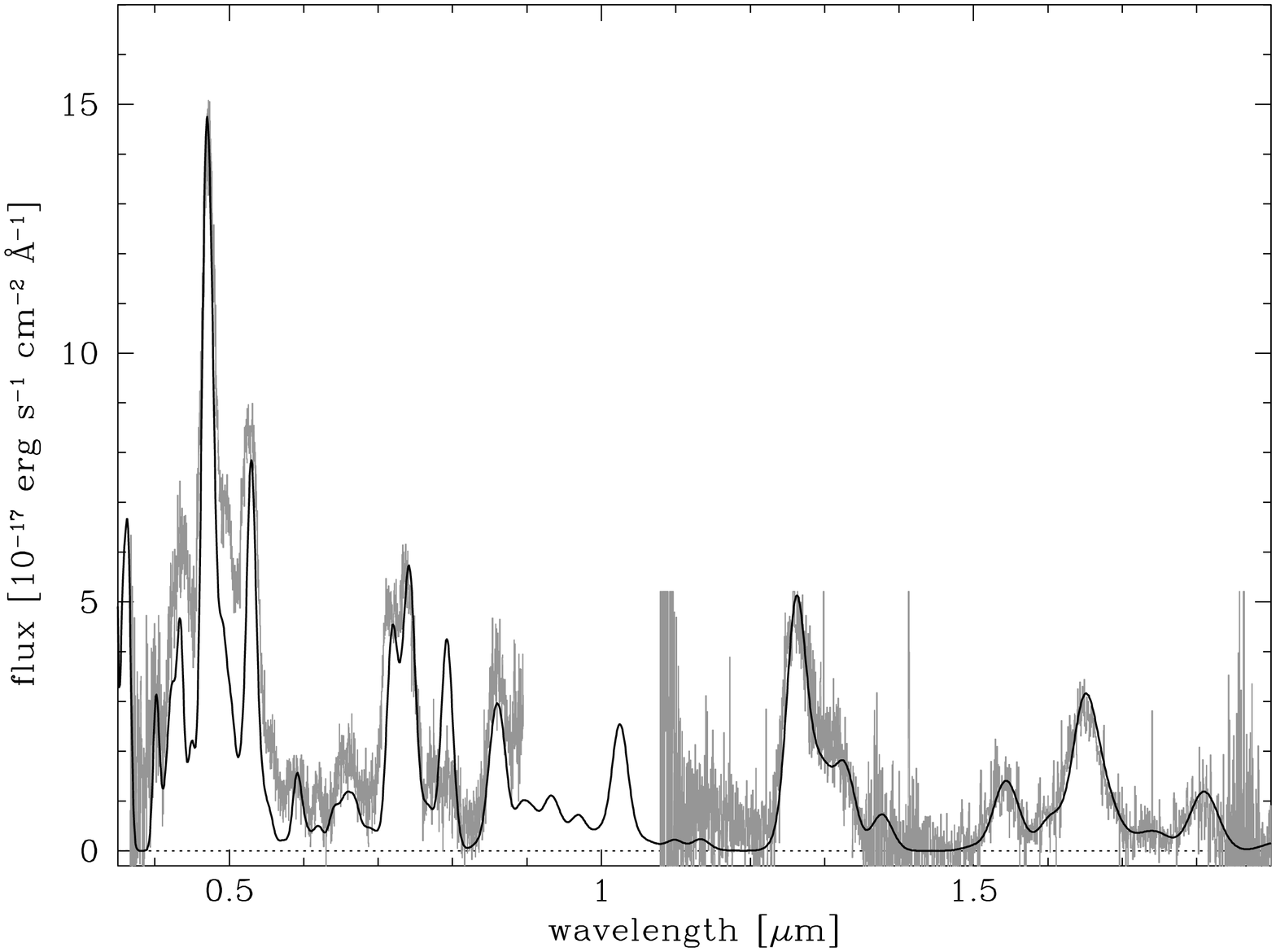}
\caption{A fit to the combined FORS1 and ISAAC spectrum using a simple
NLTE [Fe\,{\small I}], [Fe\,{\small II}] and [Fe\,{\small III}] model
and LTE [Co\,{\small II}], [Co\,{\small III}], [Ni\,{\small II}],
[Ni\,{\small III}]. The FORS1 spectrum has been scaled to match the
ISAAC one based on the photometric evolution (see text for details).
The narrow (few pixel) spikes in the data are artifacts of the
atmosphere and/or the detector.}
\label{fitisaac}
\end{figure*}

As discussed in the introduction a number of authors have used the
forbidden lines of iron in supernovae to derive masses in the ejecta.
Due to a paucity of data at epochs of one year in the near infrared
almost all studies have concentrated on the optical regime (see for
example Ruiz-Lapuente \& Lucy \cite{pilar92}; Ruiz-Lapuente et
al. \cite{pilar95}; Liu, Jeffery \& Schultz \cite{liu}). Spyromilio
et al. (\cite{spyromilio92a}) did analyse the spectra of SN\,1991T
covering a broad wavelength range. The optical regime is affected by
the reddening to a significant extent and in the absence of IR data
other authors have often also fitted the extinction. With the long
baseline provided by the IR observations the constraints on the
fitting are much stronger. However, the sensitivity of the longer
wavelength transitions to temperature and electron density are lower.
The combination of optical and IR data constrains the fit significantly.

To combine the ISAAC and FORS1 data into a single spectrum we have
first removed the echo spectrum from the FORS1 and ISAAC data
following the prescriptions discussed above. We have then increased
the flux in the FORS1 data by 0.0138 magnitudes per day (measured on
the published $V$-band light curve, excluding the echo contribution) 
making up for the 41 days of difference in the epoch of the FORS1 and 
ISAAC data.

Our modeling is based on the techniques and models described in
Spyromilio et al. (\cite{spyromilio92a}). The line identifications are
the same as those of Liu et al. (\cite{liu}) for the optical lines and
those of Spyromilio et al. (\cite{spyromilio92a}) for the infrared
lines. The temperatures and densities for our models are of order
4000\,K and 2$\times$10$^5$\,cm$^{-3}$ for the SofI epoch data and the
same temperature but lower electron density
(8$\times$10$^4$\,cm$^{-3}$) for the ISAAC epoch. For simplicity
cobalt and nickel are treated in LTE. Since we only derive abundances
from iron lines we treat these in NLTE. The atomic data are from
Nussbaumer \& Storey (\cite{nuss2}), Nussbaumer \& Storey
(\cite{nuss1}), Garstang (\cite{Garstang}), Berrington et
al. (\cite{berrington}), Grevesse, Nussbaumer \& Swings
(\cite{grevesse}), Nussbaumer \& Storey (\cite{nuss3}).

From the clean 1.644\,$\mu$m line of [Fe\,{\small II}] we can derive a
mass of iron emitting given the emissivity from the NLTE calculations
used above. Using a distance of 11\,Mpc for M96, $A_V$=0.94, and an
observed flux from the 1.644\,$\mu$m line of
1.7$\times$10$^{-14}$\,erg~s$^{-1}$~cm$^{-2}$ then the mass of singly
ionised iron present in the ejecta is derived to be 0.4\,M$_\odot$
with a conservative error of 0.1\,M$_\odot$. The
value depends solely on the emissivity for this transition. Doubly
ionized iron has strong lines in the optical. However, the
[Fe\,{\small III}] lines are heavily blended and are of higher
excitation making their emissivity less reliable. A good fit to the
[Fe\,{\small III}] lines could be achieved with a variety of
conditions. We do not believe that from our models a more accurate
mass of Fe$^{++}$ can be derived. We leave that to more sophisticated
modeling attempts in a future paper. 

Contardo, Leibundgut \& Vacca (\cite{contardo}) have analyzed the
light curves of a number of type\,Ia supernovae and find typically
0.5\,M$_\odot$ of $^{56}$Ni to have been manufactured in the
explosion. This is consistent with our result for SN\,1998bu, but
only marginally consistent with the claimed upper limit from absence of
the $\gamma$-ray line emission (Georgii et al. \cite{georgii}, 
 Milne et al. \cite{milne}).

Fransson, Houck \& Kozma (\cite{fransson}) predicted from theoretical
modeling that the near-infrared lightcurves would flatten out a year
after explosion.  As discussed earlier we observe this effect. The mass
of iron we observe emitting is inconsistent with this being the
classical infrared catastrophe as described by Axelrod
(\cite{axelrod}) whereby the flattening of the cooling curve combined
with the exponential decline of the heating due to the radioactivity
combine to produce a rapid cooling of the ejecta resulting in only
ground state energy levels being populated. Mid-infrared transitions
would remain observable but higher energy lines should vanish. In
SN\,1987A the infrared catastrophe was seen without the disappearance
of the optical and near-IR lines. However, only a small fraction of
the mass of iron was emitting in these lines. In our data presented
here this is not the case.
A significant amount of iron is still emitting in the optical and near-IR.

The flattening of the light curve may be due to the gradual transition of
the emission from the optical to the near-IR as the ejecta expand and
the densities drop. The electron densities remain high enough that the
ejecta continue to be coupled to the heating source.

We note that also the type Ia SN\,2000cx displayed a constant late light
curve in the near infrared.
The flat lightcurve was observed in the J- and H-bands
between 360-480 days past maximum, and appears consistent with late time 
modeling (Sollerman, Kozma \& Lindahl \cite{sollerman};  Sollerman, 
Kozma, Lindahl et al. \cite{sollerman2}).

\subsection{Line widths and profiles}

From the fitting of the spectra we detect a change in the line widths
of the emission lines. After correction for the recession velocity of
the parent galaxy we find that a Gaussian profile fits the observed
emission well on the red side of each line. The Gaussian profiles do
not produce a good fit to the blue side of the emission lines (see
Figs. \ref{decaysofi} and \ref{decayisaac}). The supernova seems to
emit more in the blue than in the red.

Since the lines we are observing are forbidden and moreover, are
emitted in a rapidly expanding medium, self absorption is not
considered. However, a skewed distribution of the emitting material
could easily explain the line profiles. 

The velocities required for the fits were 5000$\pm$200\,km\,s$^{-1}$
for the SofI data and 7000$\pm$200\,km\,s$^{-1}$ for the ISAAC
data. The errors on the velocities are conservative as multiple lines
are fit. This increase in the observed emitting volume implies that
the outer regions of the ejecta increased their relative emissivity
with respect to the centre between the two epochs.  The excess
emission in the blue part of the profile means that the width of the
Gaussian is constrained by the red side of the line and the expansion
velocity is a minimum. A broader skewed line profile would provide a
better fit. Since all 3 instruments were
wavelength calibrated independently we exclude a calibration error as
the source of the effect seen.

The change of the expansion velocities observed suggests
that the results of Mazzali et al. (\cite{mazzali}) relating the
observed late-time expansion velocity to the energy of the explosion
should be reconsidered as the measured velocity may depend on the
exact epoch of observation.

\subsection{Direct evidence for radioactive decay}

Varani et al. (\cite{varani}) demonstrated in the case of SN\,1987A
that the near IR lines of [Co\,{\small II}] and [Fe\,{\small II}] at
1.547 and 1.533\,$\mu$m, respectively, could be used to determine the
ratio of the mass of iron to cobalt emitting in the ejecta of a
supernova. The ratio of the two lines changes as radioactive cobalt
decays into stable iron. The two lines arise from the same ionization
stage and their excitation potentials are very similar. The ratio of the
two lines is very weakly dependent on temperature and it is expected
that the ionization structure of the cobalt and iron will be very
similar. Therefore the ratio of the lines is expected to directly
relate to the ratio of the masses of iron to cobalt.

\begin{figure}
\centering
\includegraphics[angle=-90,width=8cm]{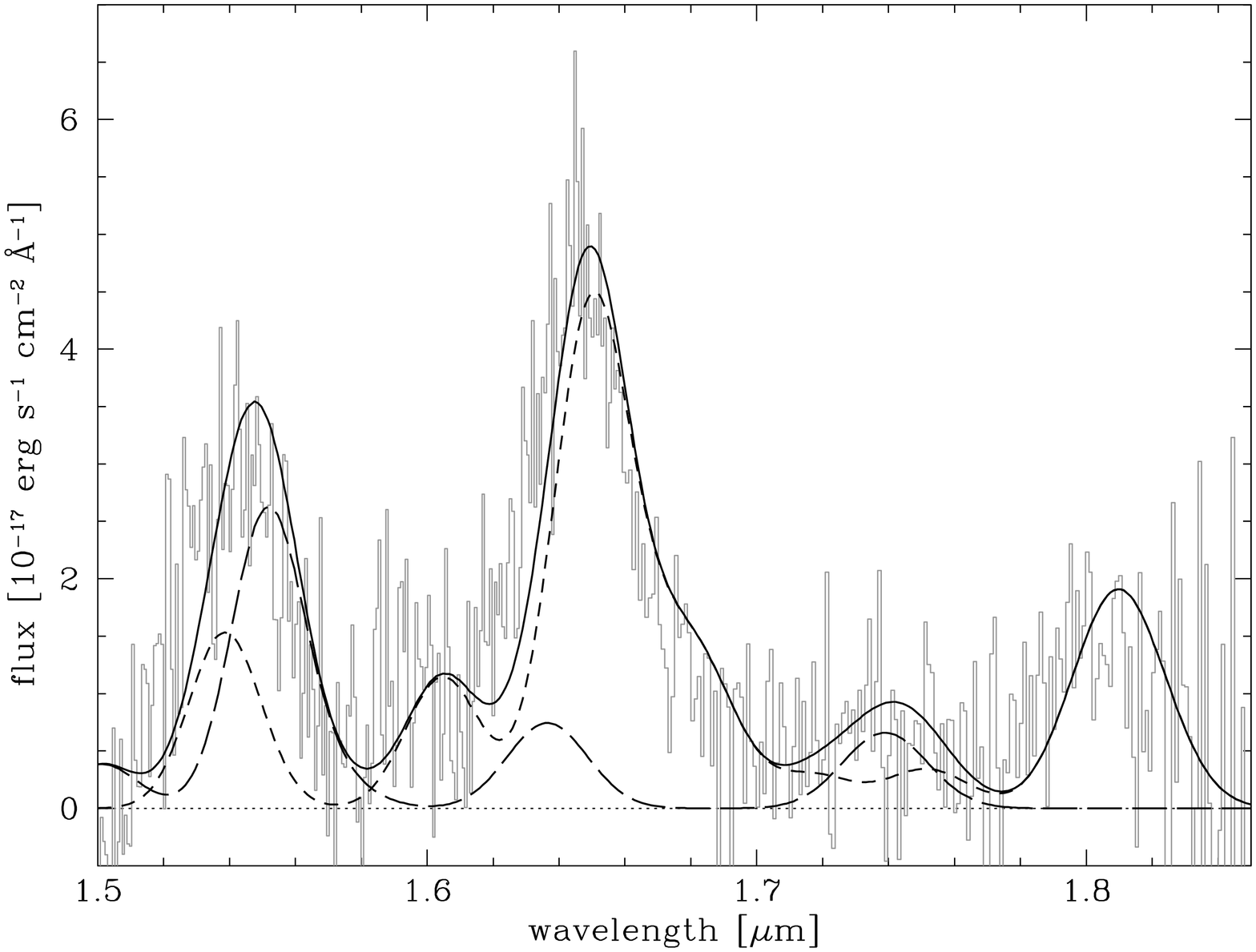}
 \caption{Same as Fig.~\ref{fitsofi} but focused on the
$H$-band. The long dashed line marks the [Fe,{\small II}] spectrum 
and the short dashed line the [Co\,{\small II}]. The continuous 
line is the sum. The 1.55\,-$\mu$m feature is fit by the 1.533
[Fe\,{\small II}] and 1.547 [Co\,{\small II}] lines.}
\label{decaysofi}
\end{figure}

\begin{figure}
\centering
\includegraphics[angle=-90,width=8cm]{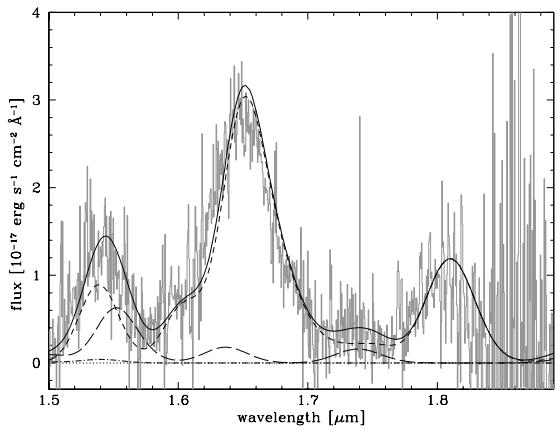}
 \caption{Same as Fig.~\ref{fitisaac} but focused on the
$H$-band. The spectra identification is as per Fig.~\ref{fitsofi}. 
Additionally the dot-dash line marks the [Fe\,{\small I}] spectrum. 
The 1.55\,-$\mu$m feature is fit by the 1.533 [Fe\,{\small II}]
and 1.547 [Co\,{\small II}] lines.}
\label{decayisaac}
\end{figure}

\begin{figure}
\centering
\includegraphics[angle=-90,width=8cm]{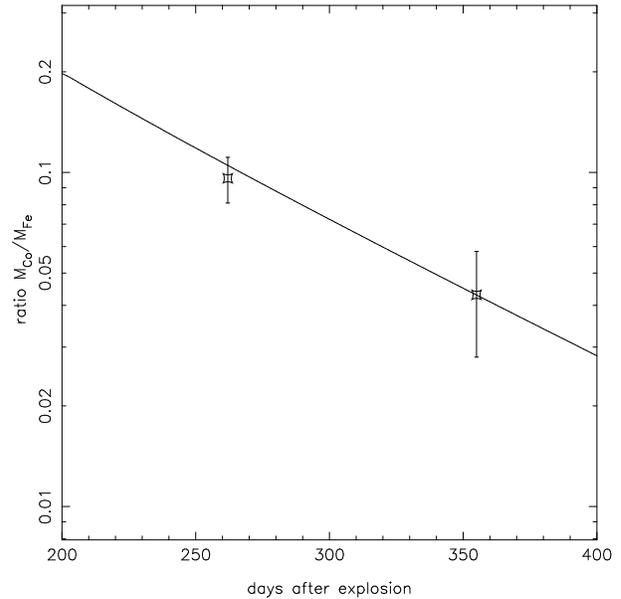}
 \caption{The derived mass ratio of Co to Fe in the ejecta of
SN\,1998bu as a function of time after explosion. The line represents
the theoretical evolution from radioactive decay and is not fit to the
data. The error bars reflect a error of 1000\,K in the assumed
temperature for the electron gas.}
\label{decay}
\end{figure}

The following formula was derived by Varani et al. (\cite{varani})
for the mass fraction:

\begin{equation}
\frac{M_{Co}}{M_{Fe}}  = \frac{F_{1.547}}{F_{1.533}} \times 0.0376
\times e^{\frac{2127}{T}}
\end{equation}

where T is the temperature of the gas in degrees Kelvin (4000\,K in our case).
 From the fits to the data (see Figs.
\ref{decaysofi} and \ref{decayisaac}) we derive the ratios and plot
them as a function of age in Fig.~\ref{decay}.

Superimposed on the two data points is the change of the mass ratio of
cobalt to iron as expected from the radioactive decay.  A pure
$^{56}$Ni initial mass is assumed for this plot with no contribution
by $^{57}$Ni. The optical cobalt and iron lines have been used in the
past by other authors (see Kuchner et al. \cite{kuchner}) to derive
similar results for type\,Ia supernovae.

\section{Conclusions}

We have used late time optical and infrared photometry and
spectroscopy of the type\,Ia
supernova 1998bu to determine the physical conditions in the ejecta a
year after the explosion. The mass of emitting iron is consistent with
determinations by other methods and in agreement with theoretical predictions.
We find that the supernova is fading very slowly in the H-band one year
after the explosion.

We find that the emission lines originating within the
ejecta appear broader and skewed at later times possibly indicating 
an assymetric distribution of iron within the ejecta.
The widths of the emission lines also increases with time.
The emission
observed in the infrared [Fe\,{\small II}] lines is consistent with 0.4
solar masses of Fe$^+$ being present in the ejecta. The evolution of
the spectrum in the near infrared shows direct evidence of the
radioactive decay of $^{56}$Co into $^{56}$Fe.

\begin{acknowledgements}

We thank the staff at the NTT for their excellent support over many
years of observing. The ISAAC observations were obtained in service
mode and we thank the Paranal staff for doing our observing for us. This
research has made use of the NASA/IPAC Extragalactic Database (NED)
which is operated by the Jet Propulsion Laboratory, California
Institute of Technology, under contract with the National Aeronautics
and Space Administration.

\end{acknowledgements}

\end{document}